\documentstyle[11pt,newpasp,twoside,epsf]{article}
\begin{document}
\title{ Infrared Imaging Polarimetry of  Massive Star-forming Regions
}
\author{Bringfried Stecklum}
\affil{TLS Tautenburg, Sternwarte 5, D-07778 Tautenburg,
Germany}
\author{Thomas Henning, Markus Feldt}
\affil{MPIA, K\"onigstuhl 17,
D-69117 Heidelberg, Germany}
\author{Hans-Ulrich K\"aufl}
\affil{ESO, K.-Schwarzschild-Str. 2, D-85748 Garching, Germany}
\author{Sebastian Wolf}
\affil{CALTECH,  Mail Code 220-6, Pasadena, CA 91125, U.S.A.}

\begin{abstract}
Imaging polarimetry is a useful tool to reveal the 3D structure of dust
distributions and to localize embedded young
stellar objects (YSOs). 
We present
maps of the linear polarization at 2.2\,\micron{} for three ultra-compact
HII regions (UCHIIs, G192.16$-$3.82, G331.28$-$0.19, G339.88$-$1.26)
and the methanol maser source G305.21+0.21. From the 
polarization maps, we draw conclusions on the morphology of these 
objects and the presence of luminous illuminating sources.
\end{abstract}

\section{Introduction}
Compared to the specific intensity, the polarization state of light provides
additional information on the radiation field. Thus polarization measurements
of light emerging from star-forming regions 
allow to retrieve important characteristics which can hardly be obtained otherwise.
At near-infrared (NIR) 
wavelengths, scattering of light by 
dust grains is the 
primary polarization process. Single scattering may lead to high polarization
degrees p,
with p($\lambda$) indicative for the grain size
(Fischer et al. 1994). Since the vibration direction of the electric vector is
perpendicular to the scattering plane defined by the incident and scattered 
rays, polarization measurements allow to localize the
illuminating source, a technique which has been widely used for
YSOs (e.g. Tamura et al. 1990; Burkert et al. 1998; Yao et al. 2000).
Massive star-forming regions, particularly UCHIIs, harbor embedded
star clusters (e.g. Feldt et al. 1998).
Although the brightest object, i.e. the most massive star, might be
hidden by dust, its scattered radiation can reach the observer if the extinction 
is non-isotropic since light emerging from directions of small optical depth will be
scattered by grains into the line-of-sight. 
The position of the primary illuminator can be estimated using the
least-squares method of Perkins, King, \& Scarrott (1981) or the centroid technique of
Weintraub \& Kastner (1993) which was applied for our targets. The positional
uncertainty arises from measurement errors as well as from the contribution of
scattered light from fainter stars, leading to both a decrease of the net
polarization and a change of the position angle $\Theta$ of the polarization vector
compared to the case of a single
source. Further depolarization can also be caused by multiple-scattering.
By means of the NIR imaging polarimetry, we wanted to localize luminous
embedded sources, study their relation with tracers of young, massive
stars like methanol masers (e.g. Walsh et al. 1997), and look if discrete
features represent stars or just scattering peaks.

\section{Observations and data reduction}
The observations were performed with SOFI at the ESO--NTT. The Wollaston prism
produces two orthogonally polarized beams, separated by 47\arcsec{} which
match well the spatial extent of our targets. A slit mask avoids the
overlapping of the two beams. Images at the pixel
scale of 0\farcs14 were taken at five dithering positions
using either the narrow-band filter 2.195\,\micron{} or the broad-band filter Ks.
Observations at two instrument orientations
yielded frames at polarization angles of 0, 45, 90 and 135\deg{}.
The observing
conditions were fine, with a seeing of $\sim$0\farcs6 and good transmission.
The measurement of G339.88$-$1.26 was carried out in June 1998 while the 
other targets were observed in March 1999. 
The data processing included sky subtraction, flat-fielding
using dome flats, and bad-pixel-correction. The target
area was extracted for each polarization angle and the resulting four images
were registered using stars in common. From the Stokes U and Q images
the quantities p and $\Theta$ were
calculated according to the standard formulae (e.g. Fischer, Stecklum, \& Leinert 1998),
taking de-biasing into account (Wardle \& Kronberg 1974).
The polarization images were rebinned to a pixel size of 0\farcs8 to
improve the signal-to-noise ratio (SNR).
The astrometry was established on the full-field total
intensity I frames based on either DSS or 2MASS and then transformed to
the stack of smaller polarization frames. The astrometric error amounts to 
$\sim$0\farcs2.
For the estimation of the location of the illuminator only
polarization vectors with $\rm p>20\%$ (predominantly single scattering)
in areas with $\rm I > 2\sigma$  were used. The error of this location is marked by the ellipse
in the figures below.

\section{Results}

\subsection{G192.16$-$3.82}
This UCHII (IRAS 05553+1631) is caused by an early B star which is
surrounded by an accretion disk (Shepherd et al. 2001).
It drives a molecular outflow (Shepherd et
al. 1998) oriented east-west. The polarization map (Fig.\,1 left) shows a bipolar structure
caused by scattering in the outflow cavities.
The brightest NIR source is a scattering peak in the eastern cavity
(blue-shifted outflow lobe) which is inclined toward the observer. Weaker
scattered light is also detected from the western lobe. The centro-symmetry of
the polarization pattern is distorted north of the brightest NIR source,
presumably due to multiple scattering. The error ellipse includes both the
UCHII and the peak of the 7\,mm emission, confirming that the embedded
object gives rise to these phenomena and the scattered light. A weak
2.2\,\micron{} source is detected at the reference position which might be the central
star, barely seen along the rim of the eastern outflow cavity. The resolution
of our images is not sufficient to prove whether ot nor it is a binary as
suggested by Shepherd et al. (2001).

\begin{figure}
\plottwo{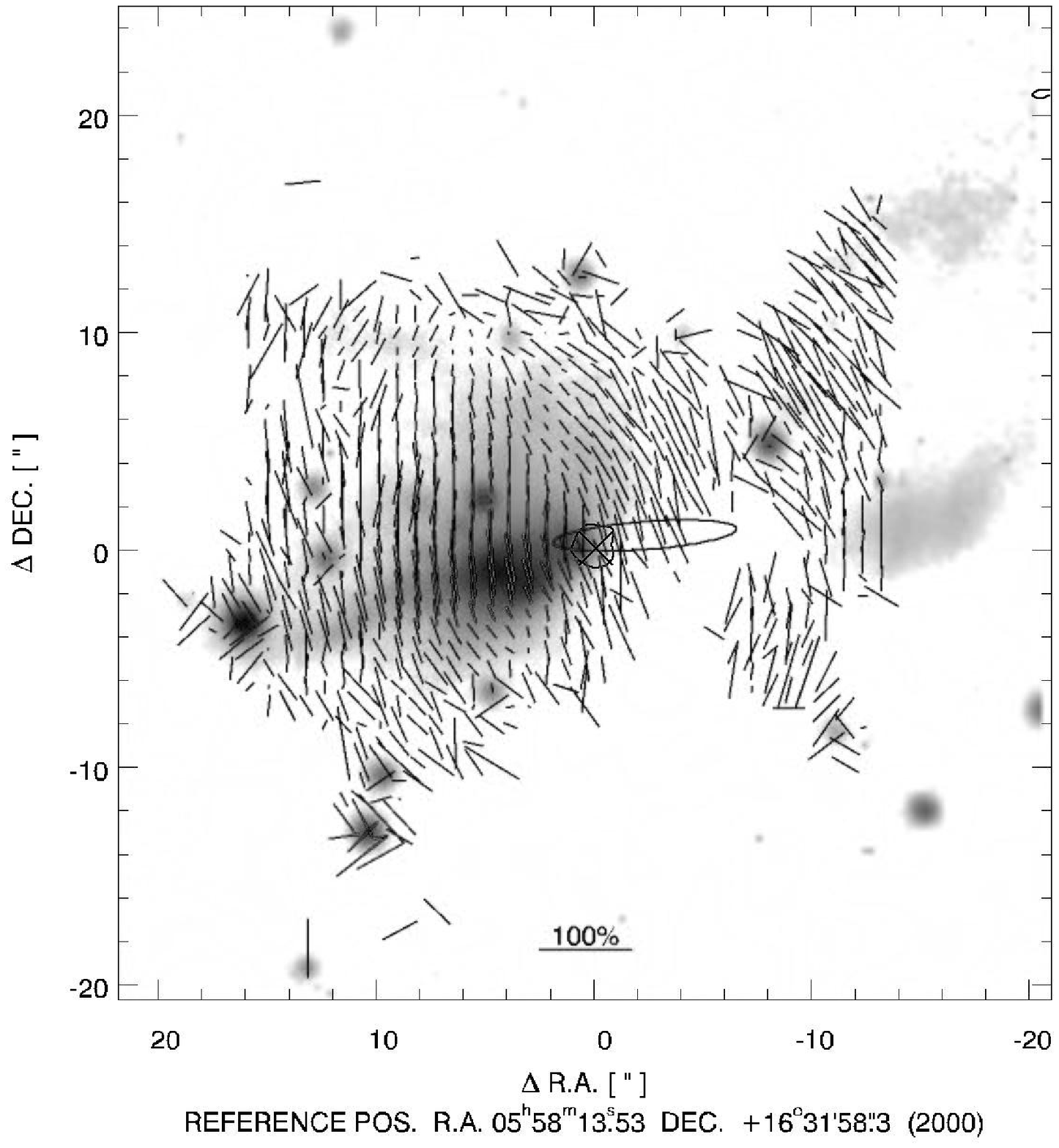}{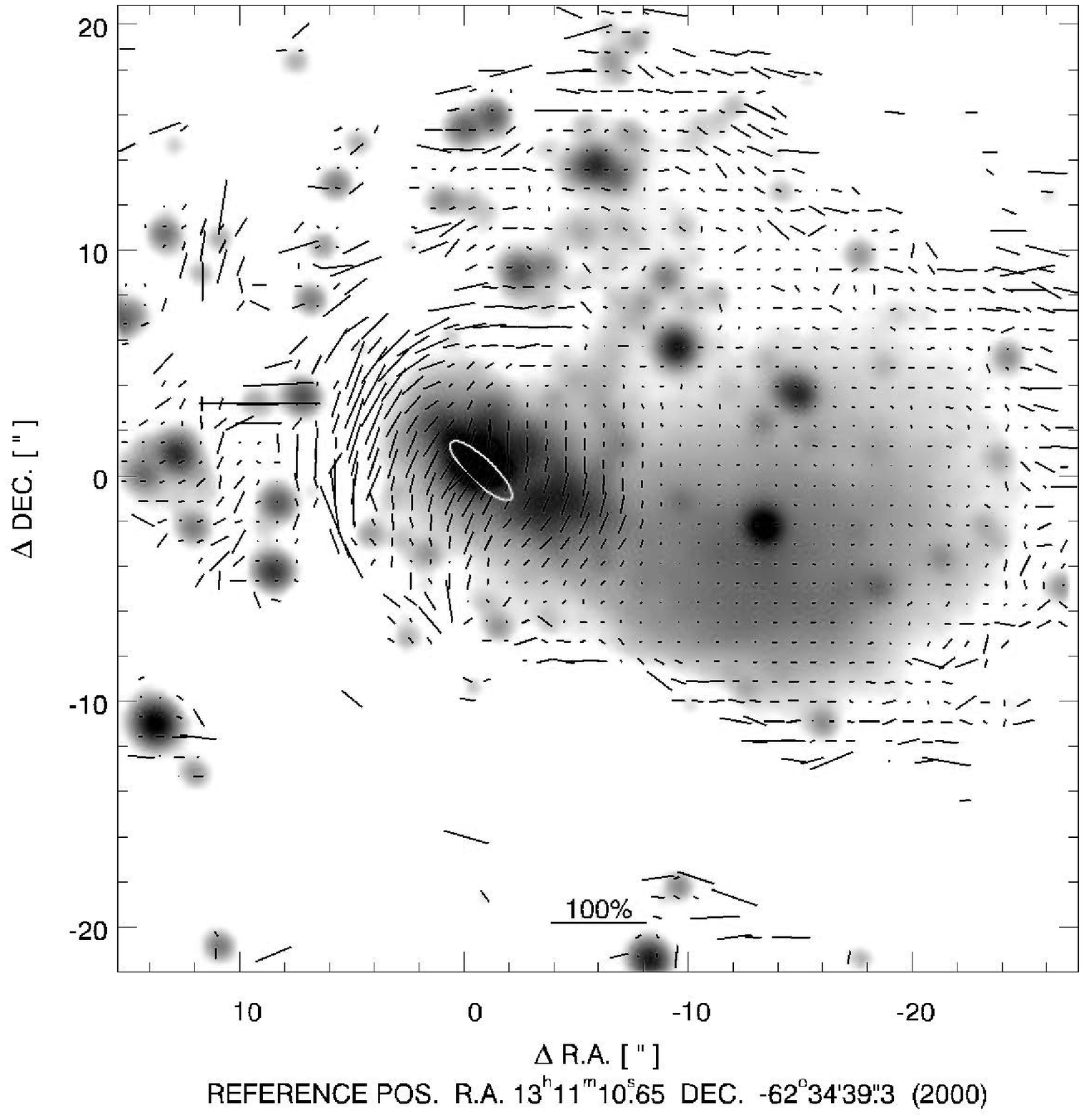}
\caption{Left: 2.195\,\micron{} image of G192.16$-$3.82 with superimposed polarization map.
The horizontal bar marks 100\% polarization. The reference position (cross) corresponds to the peak of
the 7\,mm emission (Shepherd et al. 2001).  The black contour delineates the 25\% peak level of
the 3.5\,cm radio continuum (Kurtz 1995). Right: same for G305.21+0.21. The reference position
corresponds to that of methanol masers of G305.202+0.207 (Norris et al.
1993).}
\end{figure}

\subsection{G305.21+0.21}
This object (IRAS 13079$-$6218) harbors two centers of 6.7\,GHz methanol maser
emission. While the eastern one is not detected at IR wavelengths, the western one
(G305.202+0.207, Fig.1 right) has a strong IR counterpart (Walsh et al. 2001).
Strong polarization is present in
the vicinity of the masers, indicating illumination from a compact source. At the
presumed distance of 6.2\,kpc (Philips et al. 1998), the projected separation
between the illuminator and the masers amounts to ~3000\,AU. The upper
limit for its radio continuum emission corresponds to a spectral type later than
B0.5 (Phillips et al. 1998). The MIR-to-radio flux ratio of the source exceeds
that of normal star clusters, suggesting the presence of a ``quenched'' HII
region (Walsh et al. 2001).  The kinematics of the masers were interpreted in
terms of a circumstellar disk (Phillips et al. 1998). The weak extended
emission to the southwest coincides with a small HII region (Phillips et al. 1998)
and is not strongly polarized. This might be due to diffuse illumination or a lack
of dust grains. However dust cannot be completely absent since MIR radiation
from this area has been observed (Walsh et al. 2001).

\subsection{G331.28$-$0.19}
This methanol maser source (Fig.\,2 left) has a 10\,\micron{} counterpart
(Walsh et al. 2001). The associated 8.6\,GHz radio continuum emission (Phillips et al. 1998)
peaks ~3\arcsec{} south and points to the presence
of a B0.5 star. At 2.2\,\micron{} a conical reflection nebula with a
well-defined polarization pattern is obvious. The location of the illuminator
coincides with that of the MIR source but is offset from the maser spots. The
nebula is associated with H$_2$(1$-$0)S(1) emission, shifted in
velocity with respect to the maser (Lee et al. 1999). A possible origin is a
molecular outflow although fluorescent excitation
cannot be ruled out. These features suggest that the object might be one of the youngest
massive YSOs.
\begin{figure}
\plottwo{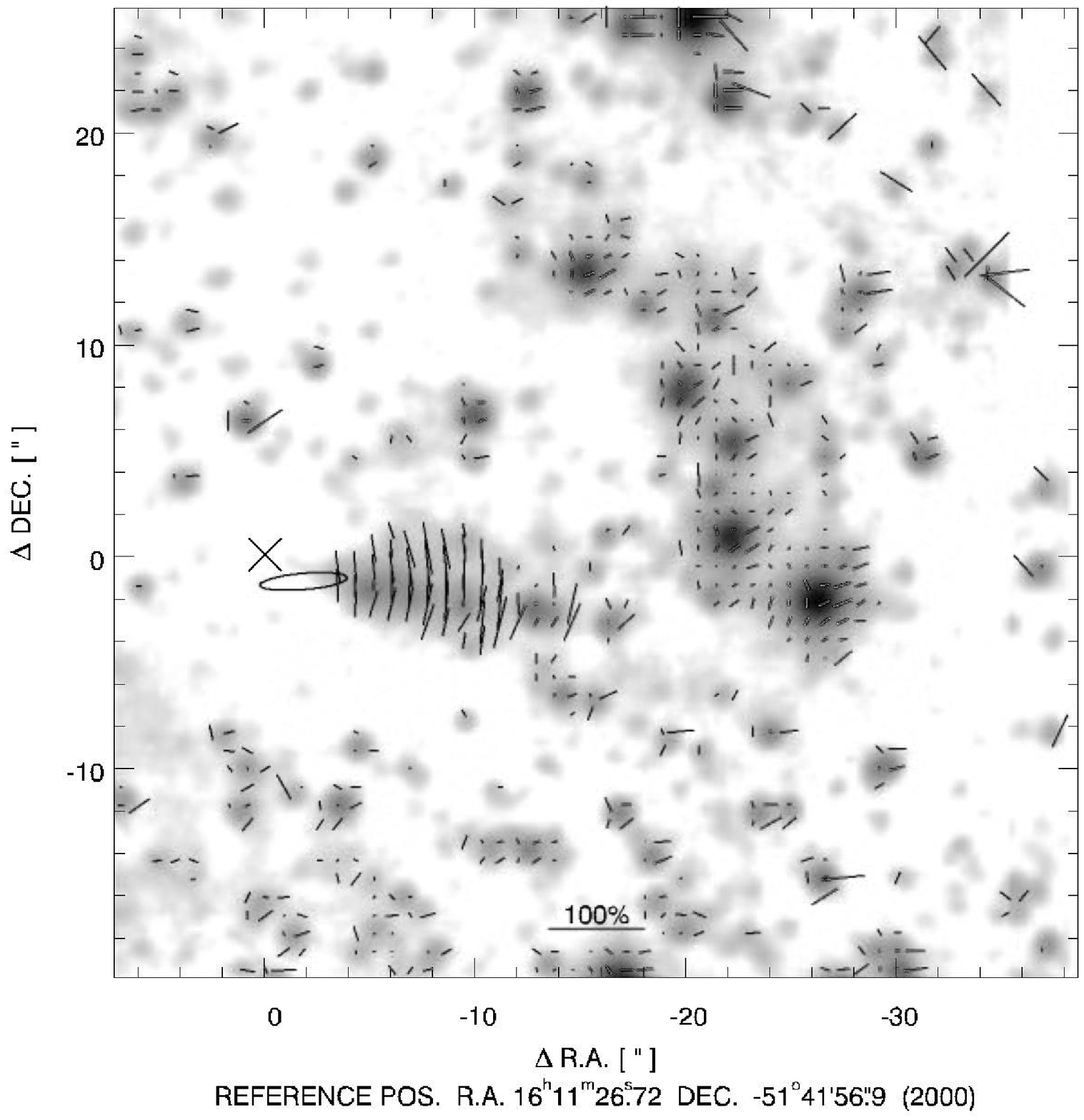}{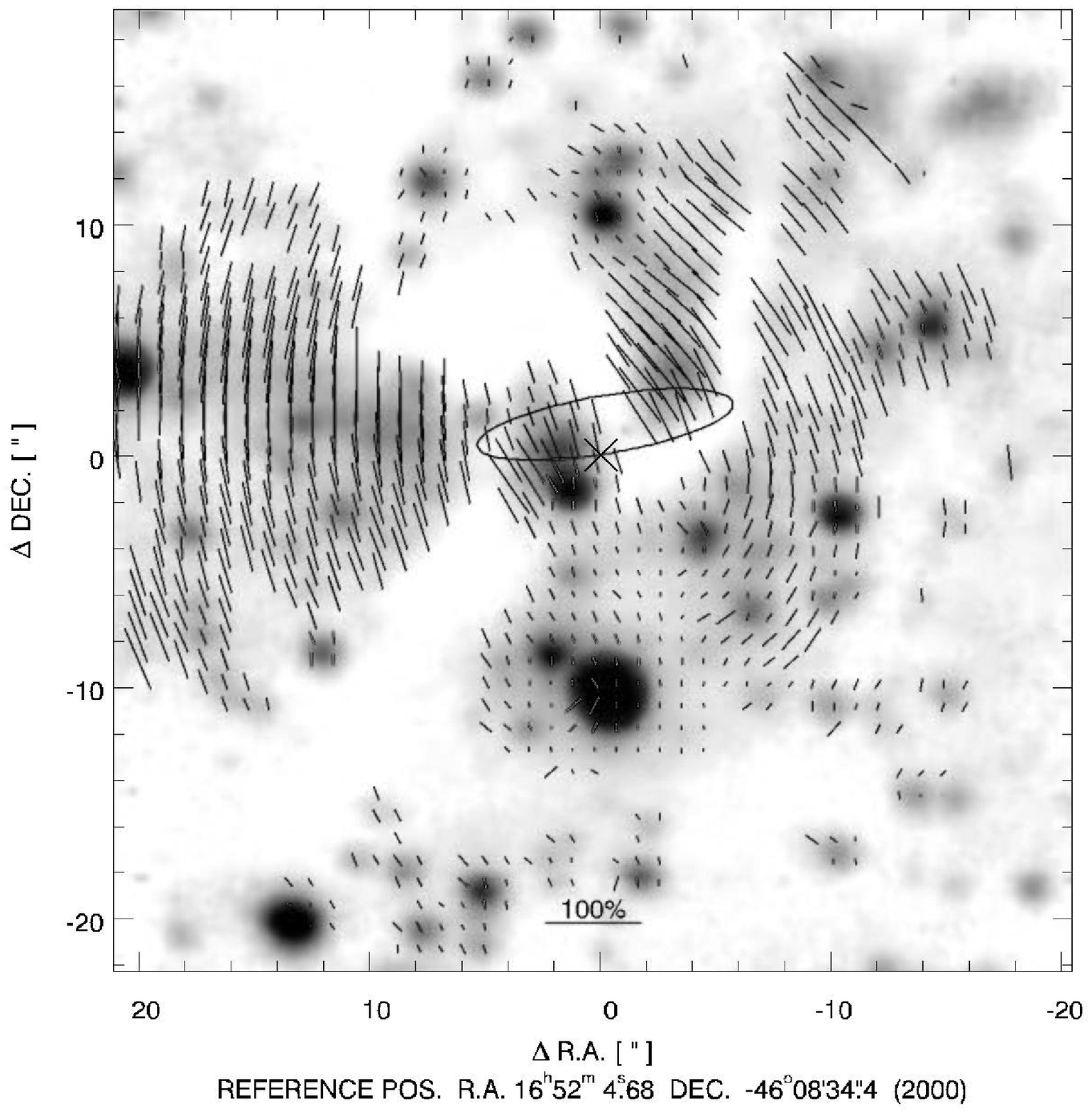}
\caption{Left: 2.195\,\micron{} image of G331.28$-$0.19 with superimposed
polarization map. Right: Ks image and polarization map of G339.88$-$1.26. For both objects the
reference position corresponds to that of methanol masers (Norris et al.
1993).
}
\end{figure}
\subsection{G339.88$-$1.26}
This UCHII harbors a chain of 6.7\,GHz methanol masers. The geometry and
velocity gradient of the masers match the circumstellar disk
model (e.g. Norris et al. 1998). Our 10\,\micron{} imaging seemed to support this
view since it showed an elliptical object well aligned with the
masers (Stecklum et al. 1998). De Buizer et al. (2000) arrived at a similar
conclusion. Thus, the source was among the prime disk candidates. 
The polarization pattern (Fig.\,2 right), however, is
inconsistent with the presumed disk orientation. In those regions which would be
shadowed by the hypothetical disk, highly polarized, nebulous emission is
present.
Our polarization data indicate that the star close to the maser
location is not the main source as claimed by De Buizer et al. (2002)
but a foreground object.
The primary illuminator  is situated behind a dust filament and thus hidden at
NIR wavelengths.
It seems likely that it corresponds to the object 1B of De Buizer et al. (2002).

\section{Conclusions}
Our NIR imaging polarimetry of four massive star-forming regions yielded
maps containing partial or almost complete centro-symmetric
pattern of the polarization vectors. This implies the presence of a
compact radiation source, either a single massive star or a dense cluster.
These illuminators were detected at 2.2\,\micron{}
for two targets but no NIR counterparts were found for the other two regions. 
In these cases, the young, massive stars are heavily obscured by dust which 
may reside in a broken-up cocoon or foreground filament. The luminous
illuminating sources are very close to methanol masers proving that these 
masers are good tracers of early stages of massive star formation.

\end{document}